\documentclass[journal]{article}
\usepackage[pass,a4paper]{geometry}
\usepackage[english]{babel}
\usepackage{amsmath,epsfig,amsfonts,dsfont,amsthm}
\usepackage{cite}
\usepackage{authblk}

\newtheorem{theorem}{Theorem}

\newcommand{\bX}{\boldsymbol{X}}
\newcommand{\bx}{\boldsymbol{x}}
\newcommand{\by}{\boldsymbol{y}}
\newcommand{\ba}{\boldsymbol{a}}
\newcommand{\bA}{\boldsymbol{A}}
\newcommand{\bB}{\boldsymbol{B}}
\newcommand{\bD}{\boldsymbol{D}}

\newcommand{\bL}{\boldsymbol{L}}

\newcommand{\bP}{\boldsymbol{P}}
\newcommand{\bW}{\boldsymbol{W}}
\newcommand{\bR}{\boldsymbol{R}}

\newcommand{\bu}{\boldsymbol{u}}
\newcommand{\bv}{\boldsymbol{v}}
\newcommand{\bU}{\boldsymbol{U}}
\newcommand{\bV}{\boldsymbol{V}}
\newcommand{\bh}{\boldsymbol{h}}
\newcommand{\bY}{\boldsymbol{Y}}
\newcommand{\bZ}{\boldsymbol{Z}}

\newcommand{\R}{\mathds{R}}

\newcommand{\bc}{\boldsymbol{c}}

\title {A mathematical framework for amplitude and phase noise analysis of coupled oscillators}

\author[1]{M. Bonnin \thanks{michele.bonnin@polito.it}}
\author[1]{F. Corinto}
\author[2]{V. Lanza }
\affil[1]{Department of Electronics and telecommunications, Politecnico di Torino}
\affil[2]{Normandie Univ, France; ULH, LMAH}

\begin{document}
\maketitle

%
\abstract{
Synchronization of coupled oscillators is a paradigm for complexity in many areas of science and engineering. Any realistic network model should include noise effects. We present a description in terms of phase and amplitude deviation for nonlinear oscillators coupled together through noisy interactions. In particular, the coupling is assumed to be modulated by white Gaussian noise. The equations derived for the amplitude deviation and the phase are rigorous, and their validity is not limited to the weak noise limit. We show that using Floquet theory, a partial decoupling between the amplitude and the phase is obtained. The decoupling can be exploited to describe the oscillator's dynamics solely by the phase variable. We discuss to what extend the reduced model is appropriate and some implications on the role of noise on the frequency of the oscillators.}
%
%
%
\section{Introduction} \label{intro}

Periodically driven oscillators and coupled oscillators are classical problems in nonlinear dynamics, with many relevant applications in physics, chemistry, biology and engineering \cite{kuramoto1984,pikovsky2001,izhikevich2006}. To make the models more realistic, external inputs can be included, to represent the unavoidable random fluctuations that
occur in real world systems, due to the physical properties of the oscillators or induced by the environment. Such disturbances can be modeled by stochastic forces applied to the oscillators, which are then described by stochastic differential equations \cite{gardiner1985,oksendal2003}.

Corrupting noise can dramatically affect the performance of oscillators. This is of particular relevance, for instance, in the field of modern electronic devices. Phase noise in oscillators can produce distortion or complete loss of incoming information in traditional receivers, and high bit error rates in phase modulated applications. Traditionally, the action of noise on electronic oscillators has been described as purely diffusive process \cite{kartner1990,demir2000}. It is commonly assumed that the effect of white noise on the spectrum of an oscillator is to produce a broadening of the oscillator's spectrum without affecting the positions of the peaks. Recently, this assumption has been questioned by the analysis of some simple solvable models, and by the development of improved mathematical techniques \cite{yoshimura2008,bonnin2013,bonnin2014}. These works have shown that the phase noise problem is best described as a convection--diffusion process, i.e. white noise may also be responsible for a shift in the oscillator's angular frequency.

It may sound surprising that a random perturbation can produce some kind of coherent modification to the oscillator's frequency. In fact one may expect that, as a result of their random nature, fluctuations have a null net effect and leave the oscillation frequency and amplitude unaffected. Random perturbations may produce a coherent modification to the oscillator's frequency because of the peculiar characteristics of oscillators. First, the autonomous nature of oscillators implies that any time shifted version of a solution is still a solution. The consequence is that phase shifts are not absorbed, but rather they accumulate in time. Second, oscillators are nonlinear dynamical systems. Some directions are preferred to others, so that perturbations along some directions are amplified, while others are attenuated. The result is that coherent behavior can emerge from random excitations.

In the last few years the idea on the role of noise has also changed at a more fundamental level. For long time noise has been considered a nuisance to be reduced as much as possible. Only recently it was figured out that noise can play a constructive role in natural phenomena or engineering applications. Important examples are stochastic resonance, where a periodic signal is amplified by noise \cite{benzi1982,wiesenfeld1995,gammaitoni1998}, and energy harvesting, where noise is used as a power source \cite{beeby2006,gammaitoni2012}. It also been recognized that in particular situations, noise can favor the synchronization of oscillators. Synchronization is the result of two competing mechanisms. On the one hand, differences in the oscillators free running frequencies are destructive to synchronization. On the other hand, couplings between the oscillators favor the emergence of collective rhythms. Similarly, can at the same work both towards and against synchronization. The phase diffusion produced by noise obviously acts against the synchronicity, while the frequency shift produced by noise may decrease the frequency mismatch between oscillators thus enhancing the emergence of locked states \cite{hauschildt2006,nakao2007}.

To study the balance between the phase diffusion and the frequency shift, one needs a mathematical model capable to capture the influence of noise on the phase of the oscillators. In this paper we present a mathematical framework to reduce a network of oscillators subject to white Gaussian noise described in terms of state variables to the equivalent amplitude and phase model. In section \ref{noisy network} we introduce some basic concepts about noisy oscillators and the theory of stochastic differential equations, that represents the ideal mathematical framework for the analysis of such problems. In section \ref{amplitude phase equations} we derive the main result of the paper, giving a rigorous description in terms of amplitude and phase variables for a network composed by nonlinear coupled oscillators of any order, subject to white Gaussian noise. We also show how the phase dynamics can be partially decoupled from the amplitude dynamics, thus suggesting the possibility to derive reduced order model analogous to the celebrated Kuramoto model \cite{kuramoto1984}. In section \ref{discussion} we discuss some implications of our model on the phase dynamics, with particular attention to the role of the noise on the expected frequencies of the oscillators. We discuss the physical origin of the frequency shift and why such an effect should not be neglected with respect to the phase diffusion process. In section \ref{example} we present a simple example to show the application of the transformation to amplitude and phase variable. The example chosen admits an analytical solution to illustrate the influence of noise on a small  network. Section \ref{conclusions} is devoted to conclusions.

\section{Noisy oscillators and stochastic differential equations}\label{noisy network}
Nonlinear oscillators can be conveniently described by the differential equation
\begin{equation}\label{eq1}
\dfrac{d \bX(t)}{dt} = \ba(\bX(t),\boldsymbol{\xi}(t))
\end{equation}
where $\bx: \R \mapsto \R^n$ is the state of the oscillator, and $\boldsymbol{\xi}$ represents the unavoidable noise sources, both internal and external, that affect the oscillator. In most practical situation the noise level is expected to be small with respect to the oscillator state, so that we are legitimate to linearize equation \eqref{eq1} around the noiseless state
\begin{equation}\label{eq2}
\dfrac{d \bX(t)}{dt} =  \ba(\bX(t),0) + \dfrac{\partial \ba(\bX(t),0)}{\partial \boldsymbol \xi} \, \boldsymbol{\xi}(t) + \ldots
\end{equation}
Another common assumption is that the noise possesses some ``nice'' statistical properties. In particular we shall assume that any noise source can be modeled as a Gaussian white noise. This assumption is justified in a wide range of practical situations, e.g. molecular dynamics, thermal noise, shot noise and Johnson noise in electronics. In general, white noise is a good approximation to a colored noise process in the case where the typical time scales of the underlying deterministic dynamics are much smaller than the noise correlation time (quasi--white approximation). By the central limit theorem, it is reasonable to describe $\boldsymbol{\xi}(t)$ as Gaussian distributed. Equation \eqref{eq2} can be rewritten as the stochastic differential equation (SDE) \cite{oksendal2003,gardiner1985}
\begin{equation}\label{eq3}
d \bX = \ba(\bX) \, d t+ \varepsilon \bB(\bX) \, d \bW
\end{equation}
where $\bX: \R \mapsto \R^n$ is the state of the oscillator, $\ba: \R^n \mapsto \R^n$ is a vector field that describes the oscillator dynamics, $\bB : \R^n \mapsto \R^{n,m}$ is a modulating real valued matrix, and $\varepsilon$ is a parameter that measures the noise intensity. $\bW : \R \mapsto \R^m$ is a vector of Wiener processes, also called Brownian motions, a continuous time stochastic process characterized by zero expectation value $E[\bW]=0$, independent increments, and with a Gaussian distribution. The vector valued function $\ba(\bX)$ is called the drift term, while the matrix $\bB(\bX)$ is called diffusion the diffusion term. For matrices $\bB$ with constant entries the noise is said additive, while for a state dependent matrices $\bB(\bX)$ the noise is said multiplicative.

Stochastic processes are nowhere differentiable, consequently the SDE \eqref{eq3} should be interpreted as a shorthanded version of the integral equation
\begin{equation}\label{eq4}
\bX(t) =  \bX(0) + \int_0^t \ba(\bX(s))  ds + \varepsilon \int_0^t \bB(\bX(s)) d\bW(s)
\end{equation}
The first integral on the right hand side is a Riemann integral, and it does not pose any particular problem. The second integral is a Riemann--Stieltjes type integral, but differently from common Riemann--Stieltjes integrals, the point at which the function is evaluated do matter. The two main interpretation schemes are Stra\-to\-no\-vich and It\^o. According to Stratonovich, the stochastic integral\footnote{Conforming to the standard notation, we use the symbol $\bB(\bX)\circ d\bW$ to denote Stra\-to\-no\-vich integral, while we reserve notation $\bB(\bX)d\bW$ for It\^o integral.} $\int_0^t B(X(s)) \circ dW(s)$ is defined as the mean square limit of the partial sum \cite{oksendal2003}
\[ S_n^S = \sum_{i=1}^n B \left( \frac{X(t_i) + X(t_{i-1})}{2} \right) \left[W(t_{i})-W(t_{i-1}) \right] \]
By contrast, in It\^o interpretation the stochastic integral $\int_0^t G(s) dW(s)$ is the mean square limit of \cite{oksendal2003}
\[ S_n^I = \sum_{i=1}^n G(t_{i-1}) \left[W(t_{i})-W(t_{i-1}) \right]\]
Both interpretations have their own pros and cons. The main advantage of Stratonovich interpretations is that traditional calculus rules apply. The drawback is that in each time interval, both the initial value $X(t_{i-1})$ and the final value $X(t_i)$ of the stochastic process $X$ are required to solve the SDE. This feature is known as the ``look in the future property'' of Stratonovich integral. As a consequence the Stratonovich interpretation is not well suited for numerical integration schemes. Moreover, in Stratonovich interpretation the stochastic process and the noise increments are correlated, making the determination of stochastic expectations difficult. By contrast, It\^o stochastic integral only requires the initial value of the stochastic process in each time interval. Therefore It\^o interpretation is preferred in the implementation of numerical integration schemes. In It\^o SDEs the stochastic process and the noise increments are uncorrelated, making the determination of stochastic expectations easier. The drawback of It\^o view is that a new set of calculus rules, known as It\^o calculus must be used.

The relevant consequence of the two different interpretations is that the same SDE has different solutions whether it is interpreted following Stratonovich or It\^o. However, the two interpretations are linked by a transformation that converts any Stratonovich (respectively It\^o) SDE into an equivalent It\^o (respectively Stratonovich) SDE. By equivalent we mean a different SDE, interpreted with different rules, but that has the same solution \cite{gardiner1985,oksendal2003}. The equivalence opens the possibility to switch from one interpretation to the other to take advantage of the pros of both the definitions. The Stratonovich SDE (the apex $S$ and $I$ denote Stratonovich and It\^o, respectively)
\begin{equation}\label{eq5}
d \bX = \ba^S(\bX)~d t + \varepsilon \bB^S(\bX) \circ d\bW
\end{equation}
is equivalent to the It\^o SDE
\begin{equation}\label{eq6}
d \bX = \ba^I(\bX)~d t + \varepsilon \bB^I(\bX) \, d\bW
\end{equation}
where ($a_i$ is the $i$-th component of $\ba$, $B_{ij}$ the $(i,j)$ element of $\bB$)
\begin{equation}\label{eq7}
a_i^I(\bX)= a_i^S(\bX) + \dfrac{\varepsilon^2}{2} \sum_{j,k} \dfrac{\partial B_{ij}}{\partial x_k} \, B_{jk}
\end{equation}
and
\begin{equation}
\bB^S(\bX) = \bB^I(\bX)
\end{equation}
In the following we shall use It\^o interpretation, omitting the apex $I$ for simplicity.

\section{Amplitude and phase equations for a network of oscillators with noisy interactions}\label{amplitude phase equations}

A network composed of $N$ weakly coupled nonlinear oscillators with noisy interactions can be described, similarly to \eqref{eq3}, by the (It\^o) SDEs
\begin{equation}
d \bX_i = \left[ \ba_i(\bX_i) + \varepsilon \bc_i(\bX_1,\ldots,\bX_N) \right] dt + \varepsilon \bB_i(\bX_1, \ldots, \bX_N) \, d\bW_i \qquad i=1,\ldots,N \label{eq9}
\end{equation}
where $\bX_i: \R \mapsto \R^{n}$ is a stochastic process describing the state of the $i$--th oscillator, $\ba_i : \R^n \mapsto \R^n$ is the $i$--th drift coefficient, $\bB_i: \R^{n\cdot N} \mapsto \R^{n,m}$ is the $i$--th $n \times m$ diffusion matrix, and $\bW_i: \R \mapsto \R^{m}$ is the $i$--th vector of Wiener processes. For the sake of simplicity, in equation \eqref{eq9} we assume that all oscillators are of the same order\footnote{Oscillators of different order do not pose any particular problem, they only make the notation more involved. The theorems \ref{theorem1} and \ref{theorem2} can be formulated, mutatis mutandis, for oscillator of different orders.} ($\bX_i \in \R^n$, for all $i$), but we allow the interaction to vary for each oscillator both in the modulating matrix $\bB_i$ and in the random fluctuation $\bW_i$. Such a model may arise, for instance, if both $\ba_i$ and $\bc_i$ have some small stochastic components, and they are linearized around a noiseless state.

For $\varepsilon = 0$, the SDE \eqref{eq9} reduce to an ordinary differential equation (ODE) describing $N$ independent, noiseless oscillators. The $i$--th oscillator is described by the ODE
\begin{equation}
\dfrac{d \bx_i(t)}{dt} = \ba_i(\bx_i(t)) \label{eq10}
\end{equation}
We assume that the ODE \eqref{eq9} admits an asymptotically stable $T_i$--periodic solution, represented by a limit cycle $\bx_{S_i}(t)$ in its state space. For each oscillator we define the vector
\begin{equation}
\bu_{1_i}(t) = \dfrac{\ba_i(\bx_{S_i}(t))}{|\ba_i(\bx_{S_i}(t))|}\label{eq11}
\end{equation}
$\bu_{1_i}(t)$ is the unit vector that at each time instant is tangent to the limit cycle $\bx_{S_i}(t)$. Together with $\bu_{1_i}(t)$ we consider other $n-1$ vectors $\bu_{2_i}(t),\ldots, \bu_{n_i}(t)$, such that the set $\{\bu_{1_i}(t),\ldots, \bu_{n_i}(t)\}$ is a basis for $\R^{n}$ for all $t$. Let $\bU_i(t) = [\bu_{1_i}(t),\ldots,\bu_{n_i}(t)]$ be the matrix whose columns are $\bu_{1_i}(t),\ldots, \bu_{n_i}(t)$. Such a matrix is obviously invertible, and let $\bV_i(t) = \bU_i^{-1}(t)$ be the inverse. We define the reciprocal vectors $\bv_{1_i}^T(t),\ldots,\bv_{n_i}^T(t)$ to be the rows of $\bV_i(t)$. By construction, the $\bu_i$ and $\bv_i$ vectors are bi--orthogonal, i.e.
\[ \bu^T_{\alpha_i}(t) \bv_{\beta_i}(t) = \delta_{\alpha\beta} \]
where $\delta_{\alpha\beta}$ is the Kronecker's symbol.  We shall also use the matrices $\bY_i(t) = [\bu_{2_i}(t),\ldots,\bu_{n_i}(t)]$, $\bZ_i(t) = [\bv_{2_i}(t),\ldots,\bv_{n_i}(t)]$, and the modulus of the vector field $r_i(t) = |\ba_i(\bx_{S_i}(t))|$ evaluated over the limit cycle.

\begin{figure}
\centering
 \includegraphics[width=60mm]{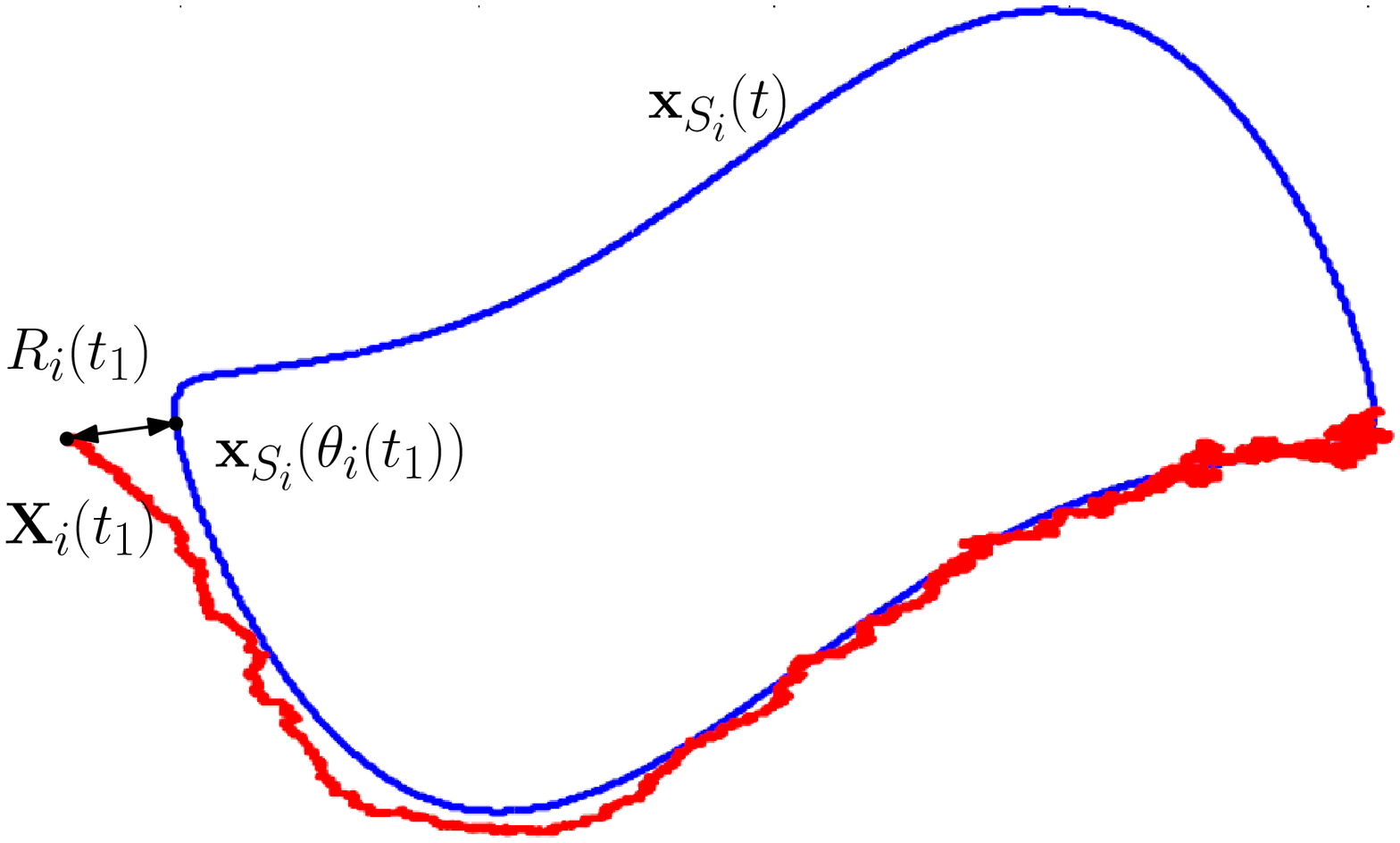}
 \includegraphics[width=60mm]{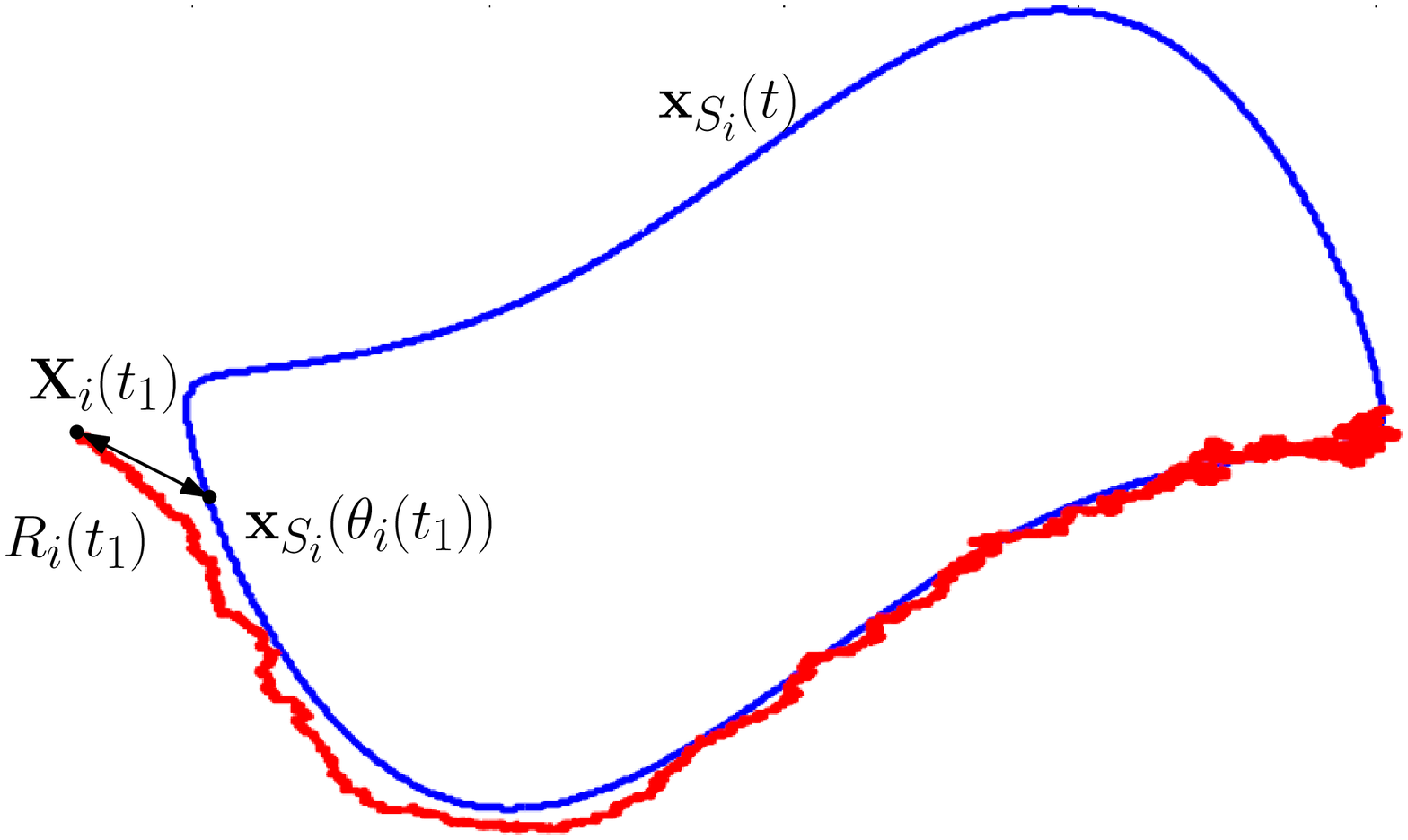}
\caption{Two possible decompositions of the stochastic process $\bX_i(t)$. At the time $t_1$ the process is decomposed as $\bX_i(t_1) = \bx_{S_i}(\theta_i(t_1)) + \bY_i(\theta_i(t_1)) R_i(t_1)$ using two different basis vectors. Left: orthogonal basis. Right: ``oblique'' basis. Red line is the stochastic process $\bX_i$, blue line is the limit cycle $\bx_{S_i}(t)$ shown for reference.}\label{figure1}
 \end{figure}

A crucial concept to be defined in the analysis of synchronization of oscillators is the phase concept. A phase function is intended to represent the projection of the oscillator's state onto a reference trajectory, normally the unperturbed limit cycle.
For each oscillator we introduce a phase function $\theta_i : \R^{n} \mapsto [0,T_i)$, interpreted as an elapsed time from an initial reference point. Consider a point $\bx_{S_i}(0)$ on the limit cycle, and assign phase zero to this point, i.e. $\theta_i(\bx_{S_i}(0)) = 0$. The phase of the point $\bx_{S_i}(t)$ is $\theta_i(\bx_{S_i}(t)) = t, \mod T_i$. Thus, the phase represents a new parametrization of the limit cycle. Together with the phase function we shall consider an amplitude deviation function $\bR_i : \R^{n} \mapsto \R^{n-1}$, with  $\theta_i, \bR_i \in \mathcal{C}^m(R^n)$, $m\ge2$. The amplitude\footnote{We shall use the term ``amplitude'' instead of the more correct ``amplitude deviation'' for the sake of simplicity.} function $\bR_i$ is interpreted as an orbital deviation from the limit cycle $\bx_{S_i}(t)$, see figure \ref{figure1}.

The following theorem establishes the amplitude and phase equation for the network
\begin{theorem}\label{theorem1}
Consider the It\^o diffusion \eqref{eq9}, and the reciprocal bases $\{\bu_{1_i}(t),\ldots,\bu_{n_i}(t)\}$ and $\{\bv_{1_i}(t),\ldots,\bv_{n_i}(t)\}$, satisfying the bi--orthogonality condition $\bv_{\alpha_i}^T \bu_{\beta_i} = \delta_{\alpha \beta}$, for all $i=1,\ldots,N$. Consider the coordinate transformation
\begin{equation}
\bx_i = \bh_i(\theta_i,\bR_i) = \bx_{S_i}(\theta_i(t)) + \bY_i(\theta_i(t)) \, \bR_i(t)\label{eq12}
\end{equation}
Then in a neighborhood of the limit cycle $\bx_{S_i}$ the phase $\theta_i(t)$ and the amplitude $\bR_i(t)$ are It\^o processes and satisfy
\begin{subequations}
\begin{align}
d\theta_i = & \big[1 + a_{\theta_i}(\theta_i,\bR_i) + \varepsilon^2 \, \hat a_{\theta_i}(\theta_1\ldots\bR_N) + \varepsilon c_{\theta_i}(\theta_1 \ldots \bR_N)\big] dt + \varepsilon \bB_{\theta_i}(\theta_1 \ldots \bR_N) \, d\bW_i \label{eq13a}\\
d \bR_i = & \big[\bL_i(\theta_i) \bR_i + \ba_{\bR_i}(\theta_i,\bR_i) + \varepsilon^2 \hat \ba_{\bR_i}(\theta_1 \ldots \bR_N) + \varepsilon \bc_{\bR_i}(\theta_1 \ldots \bR_N) \big] dt + \varepsilon \bB_{\bR_i}(\theta_1 \ldots \bR_N) \, d\bW_i \label{eq13b}
\end{align}
\end{subequations}
where $(\theta_1 \ldots \bR_N)$ is a shorthanded notation for $(\theta_1, \bR_1, \ldots, \theta_N, \bR_N)$ and (we omit explicit dependence on $\theta_i$ and $t$ for simplicity)
\begin{subequations}
\begin{align}
a_{\theta_i}(\theta_i,\bR_i) & =  \left(r_i + \bv_{1_i}^T \dfrac{\partial \bY_i}{\partial \theta_i} \bR_i \right)^{-1} \bv_{1_i}^T \bigg[\ba_i(\bx_{S_i} +\bY_i \bR_i) - \ba_i(\bx_{S_i}) - \dfrac{\partial \bY_i}{\partial \theta_i} \bR_i \bigg] \label{eq14a} \\[1ex] \nonumber
\hat a_{\theta_i}(\theta_1 \ldots \bR_N) & = - \left(r_i + \bv_{1_i}^T \dfrac{\partial \bY_i}{\partial \theta_i} \bR_i \right)^{-1} \bv_{1_i}^T  \bigg[ \dfrac{\partial \bY_i}{\partial \theta_i} \bB_{\bR_i} \bB_{\theta_i}^T \\
&  + \dfrac{1}{2} \bigg( \dfrac{\partial \ba_i(\bx_{S_i})}{\partial \theta_i} + \dfrac{\partial^2 \bY_i}{\partial \theta_i^2} \, \bR_i \bigg) \bB_{\theta_i} \bB_{\theta_i}^T \bigg]\label{eq14b} \\[1ex]
c_{\theta_i}(\theta_i,\bR_i) = & \left(r_i + \bv_{1_i}^T \dfrac{\partial \bY_i}{\partial \theta_i} \bR_i \right)^{-1} \bv_{1_i}^T  \bc_i(\bx_{S_1}+\bY_1\bR_1,\ldots,\bx_{S_N}+\bY_N\bR_N) \label{eq14c}\\[1ex]
\bB_{\theta_i}(\theta_1 \ldots \bR_N) & =  \bigg(r_i + \bv^T_{1_i} \dfrac{\partial \bY_i}{\partial \theta_i} \bR_i \bigg)^{-1}  \bv_{1_i}^T \, \bB_i(\bx_{S_1} + \bY_1 \bR_1, \ldots, \bx_{S_N} + \bY_N \bR_N) \label{eq14d}\\[1ex]
\bL_i(\theta_i) & =  - \bZ_i^T \dfrac{\partial \bY_i}{\partial \theta_i} \label{eq14e}\\[1ex]
\ba_{\bR_i}(\theta_i,\bR_i) & =  - \bZ_i^T \bigg[ \dfrac{\partial \bY_i}{\partial \theta_i} \bR_i \, a_{\theta_i} - \ba_i(\bx_{S_i} + \bY_i \bR_i) \bigg] \label{eq14f}
\end{align}
\begin{align}
\hat \ba_{\bR_i}(\theta_1\ldots\bR_N) & =   -\bZ_i^T \bigg[ \dfrac{\partial \bY_i}{\partial \theta_i} \bR_i \, \hat \ba_{\theta_i} + \dfrac{\partial \bY_i}{\partial \theta_i} \bB_{\bR_i} \bB^T_{\theta_i}  + \dfrac{1}{2} \bigg( \dfrac{\partial \ba_i(\bx_{S_i})}{\partial \theta_i} + \dfrac{\partial^2 \bY_i}{\partial \theta^2_i} \, \bR_i \bigg) \bB_{\theta_i} \bB^T_{\theta_i} \bigg] \label{eq14g} \\[1ex]
\bc_{\bR_i}(\theta_1\ldots \bR_N) & = -\bZ_i^T \dfrac{\partial \bY_i}{\partial \theta_i} \, \bR_i \, c_{\theta_i}(\bx_{S_1}+\bY_1 \bR_1,\ldots,\bx_{S_N}+\bY_N \bR_N) \label{eq14h} \\[1ex]
\bB_{\bR_i}(\theta_1\ldots\bR_N) & =  \bZ_i^T \bigg[ \bB_i(\bx_{S_1}+\bY_1 \bR_1,\ldots,\bx_{S_N} +\bY_N \bR_N)  - \dfrac{\partial \bY_i}{\partial \theta_i} \bR_i \, \bB_{\theta_i}\bigg] \label{eq14i}
\end{align}
\end{subequations}
\end{theorem}

\emph{Proof:} See appendix A.\\

The amplitude and phase equations \eqref{eq13a} and \eqref{eq13b} are exact, since no approximation is involved in their derivation, and they are valid for any value of the noise intensity $ \varepsilon$ as long as the Jacobian matrices $D\bh_i$ are regular. The amplitude and phase equations obtained crucially depend on the choice of the basis vectors $\bu_{2_i},\ldots,\bu_{n_i}$.

In general, the equations for the two It\^o processes for the phase and for the amplitude are coupled together. It is possible to show that, making use of Floquet theory, a partial decoupling between the phase and the amplitude dynamics is obtained.
Before introducing the theorem we recall the main results of the Floquet theory \cite{demir2000,bonnin2012}. Let $\bA_i(t)=\frac{\partial \ba_i(\bx_{S_i})}{\partial \bx_i}$ be the Jacobian matrix of the $i$--th oscillator evaluated on the limit cycle $\bx_{S_i}(t)$, and let $\Phi_i(t)$ be the fundamental matrix of the variational equation
\[ \dfrac{d\by_i(t)}{dt}=\bA_i \by_i(t). \]
Thus, from Floquet theory we get:
\begin{equation}\label{eq19}
\Phi_i(t)=\bP_i(t)e^{\bD t}\bP_i^{-1}(0),
\end{equation}
where $\bP_i(t)$ is a $T_i$--periodic matrix, and $\bD_i=\text{diag}[\nu_{1_i},\dots,\nu_{n_i}]$ is a diagonal matrix whose diagonal entries are the Floquet characteristic exponents \cite{demir2000,bonnin2012}.
\begin{theorem}\label{theorem2}
If the vectors $\bu_{2_i}(t),\ldots, \bu_{n_i}(t)$ are chosen such that
$$[r_i\bu_{1_i}(t),\ldots, \bu_{n_i}(t)]=\bP_i(t),$$
then the It\^o processes \eqref{eq13a} and \eqref{eq13b} become
\begin{subequations}
\begin{align}
d\theta_i = & \big[1 + \tilde a_{\theta_i}(\theta_i,\bR_i) + \varepsilon^2 \, \hat a_{\theta_i}(\theta_1\ldots\bR_N)+\varepsilon \, c_{\theta_i}(\theta_1\ldots\bR_N) \big] dt + \varepsilon \bB_{\theta_i}(\theta_1\ldots\bR_N) \, d\bW_i \label{eq20a}\\
d \bR_i = & \big[\widetilde \bD_i \bR_i+ \tilde \ba_{\bR_i}(\theta_i,\bR_i) + \varepsilon^2 \hat \ba_{\bR_i}(\theta_1\ldots\bR_N)+\varepsilon \, \bc_{\bR_i}(\theta_1\ldots\bR_N) \big] dt + \varepsilon \bB_{\bR_i}(\theta_1\ldots\bR_N) \, d\bW_i, \label{eq20b}
\end{align}
\end{subequations}
where $\widetilde \bD_i=\text{diag}[\nu_{2_i},\dots,\nu_{n_i}]$ and the Taylor series of $\tilde a_{\theta_i}(\theta_i,\bR_i)$ and $\tilde \ba_{\bR_i}(\theta_i,\bR_i)$ do not contain linear terms in $\bR_i$.
\end{theorem}

\emph{Proof:} See appendix B.\\


The asymptotic stability hypothesis of the limit cycle implies that the Floquet characteristic exponents $\nu_{2_i}, \ldots, \nu_{n_i}$ have negative real parts, for all $i=1,\ldots,N$. As a consequence, in the limit $\varepsilon \ll 1$, the amplitude asymptotic dynamics is one order of magnitude slower than the phase dynamics. This observation suggests the idea to neglect the amplitude dynamics given in \eqref{eq18b}, and to approximate the stochastic processes $\bR_i$ in \eqref{eq18a} with some reasonable (possibly constant) estimate. This approach leads to the so called ``phase reduced models''. For instance, assuming the unperturbed value $\bR_i\approx 0$, and taking into account that $a_{\theta_i}(\theta_i,0) = 0$ we obtain the simple phase equation
\begin{equation}
d \theta_i = \big(1+\varepsilon^2 \hat a_{\theta_i}(\theta_i) \big) dt + \varepsilon \bB_{\theta_i}(\theta^{(1)},\ldots,\theta^{(N)}) \, d\bW_i \label{eq21}
\end{equation}
We remark that the assumptions leading to \eqref{eq19} is justified only if the amplitudes relax instantaneously to the unperturbed value, an assumption that is often made more for mathematical convenience than being physically plausible. In fact, this assumption relies on linear approximation of manifolds, and nonlinear effects will become stronger the further we move away from the limit cycle. Moreover, the presence of nearby invariant structures such as equilibrium points and invariant manifolds may result in trajectories spending long periods of time away from the limit cycle, thus nullifying the instantaneous relaxation hypothesis. As a consequence a better solution is to chose the approximation $\bR_i\approx E[\bR_i]$, provided that the expected amplitude can be computed \cite{bonnin2014}.

\section{Discussion}\label{discussion}

A full analysis of the amplitude and phase equations \eqref{eq13a}, \eqref{eq13b} is a formidable problem and will not be treated here. We limit ourselves to some considerations and some general comments.

Equations \eqref{eq13a}, \eqref{eq13b}, or the phase reduced model \eqref{eq21} suggest that the drift effects due to $\varepsilon^2 \hat a_{\theta_i}$ become negligible in the limit of vanishing small noise ($\varepsilon \rightarrow 0$). The phase model in this limit has been extensively studied both at the single oscillator and network level \cite{kartner1990,demir2000,medvedev2010}. However, for small but finite values of $\varepsilon$, the drift effects may become significant if $\hat a_{\theta_i}$ become large enough. That the drift effects should not be neglected even for small values of $\varepsilon$ can be justified as follows. When dealing with stochastic processes, the single realization is not very much significant: it is often much more useful to look at expected quantities. To illustrate the point, it is sufficient to consider a single oscillator described by a phase reduced model, the full network model \eqref{eq13a}, \eqref{eq13b} is conceptually analogous. Let $f(\theta)$ be an arbitrary scalar function of the phase, and let $u(t,\theta)= E[f(\theta)]$ be the expected value of this function, with initial value $u(0,\theta) = f(\theta)$. Then the time evolution of $u(t,\theta)$ is governed by the Kolmogorov Backward Equation \cite{oksendal2003}
\begin{equation}
\dfrac{\partial u}{\partial t} = \mathcal{A} u, \label{eq22}
\end{equation}
where $\mathcal A$ is the generator of the It\^o diffusion
\begin{equation}
\mathcal A f(\theta) = \big(1+ \varepsilon^2 \hat a_{\theta}(\theta) \big) \dfrac{\partial f(\theta)}{\partial \theta} + \dfrac{\varepsilon^2}{2} \, \big( \bB_{\theta}(\theta) \bB_{\theta}^T(\theta)\big) \, \dfrac{\partial^2 f(\theta)}{\partial \theta^2} \label{eq23}
\end{equation}
Equation \eqref{eq23} shows that both the $\mathcal{O}(\varepsilon^2)$ drift coefficient and the $\mathcal{O}(\varepsilon)$ diffusion coefficient in \eqref{eq21} contribute for $\mathcal O(\varepsilon^2)$ terms to the evolution of expected quantities in \eqref{eq22}, \eqref{eq23}, and therefore we are not allowed to neglect one contribution with respect to the other.

Expected quantities can be determined using It\^o calculus without solving the Kolmogorov Backward Equation \eqref{eq23}. In fact, let $\bX$ be the solution of an It\^o SDE, and let $f$ be a non anticipating function (adapted process), then the zero expectation property of It\^o stochastic integral holds
\begin{equation}
E\left[\int_{t_0}^t  f(\bX) \, d\bW \right] = 0 \label{eq24}
\end{equation}
Taking the stochastic expectation on both sides of equations \eqref{eq13a}, \eqref{eq13b} and using the zero expectation property we can transform the SDEs for the amplitude and phase into a set of ODEs for the expectation values
\begin{subequations}
\begin{align}
E\left[ \frac{d \theta_i}{dt}\right] = &  1 + E\big[a_{\theta_i}(\theta_i,\bR_i)\big] + \varepsilon E[c_{\theta_i}(\theta_1\ldots \bR_N)] + \varepsilon^2 E\big[\hat a_{\theta_i}(\theta_1\ldots \bR_N)  \big]\label{eq25a} \\[2ex]
\dfrac{dE\big[\bR_i\big]}{dt}  = & E\big[\bL(\theta_i) \bR_i\big] + E\big[\ba_{2_i}(\theta_i,\bR_i) \big] + \varepsilon E[\bc_{\bR_i}(\theta_1\ldots \bR_N)] + \varepsilon^2 E\big[\hat \ba_{\bR_i}(\theta^{(1)}\ldots R^{(N)})\big] \label{eq25b}
\end{align}
\end{subequations}
where the property $dE[\theta]/dt = E[d\theta/dt]$ has been used. The problem here is the nonlinear nature of the ODEs. In fact, to compute the expectation of the nonlinear functions one needs all the moments for the amplitudes and the phases. We illustrate the issue for the function $a_{\theta_i}$ for the simple case of a scalar amplitude $R_i$. Taking the Taylor series in the neighborhood of $\theta_i=0$, $R_i=0$ we have
\begin{align}\nonumber
E\big[a_{\theta_i}(\theta_i,R_i)\big] = & a_{\theta_i}(0,0) + \dfrac{\partial a_{\theta_i}(0,0)}{\partial \theta_i} E\big[ \theta_i\big] + \dfrac{\partial a_{\theta_i}(0,0)}{\partial R_i} E\big[ R_i\big] \\
& + \dfrac{1}{2}  \dfrac{\partial^2 a_{\theta_i}(0,0)}{\partial \theta_i^2} E\big[ \theta_i^2\big] + \dfrac{1}{2}  \dfrac{\partial^2 a_{\theta_i}(0,0)}{\partial R_i^2} E\big[ R_i^2\big] + \dfrac{\partial^2 a_{\theta_i}(0,0)}{\partial R_i \partial \theta_i} E\big[ R_i \theta_i\big]\ldots \label{eq26}
\end{align}
Therefore, to compute $E[\theta_i]$ and $E[R_i]$, one needs all the moments of $\theta_i$, $R_i$, i.e. the system is open. To close the system various approaches are available \cite{hespanha2008,gillespie2009}. Among others, moment closure techniques are procedures to approximate the exact (but open) moment dynamics with a closed (but approximate) system. A relatively simple closure technique amounts to assume that higher order moments can be expressed in terms of the lowest order ones, assuming that the stochastic processes satisfy certain distribution laws (e.g. Gaussian distribution). Equation \eqref{eq26} is also instrumental to show the limit of the phase reduction method. Although the amplitude is expected to remain small ($E\big[ R_i\big] \approx 0$), higher order moments (e.g. the variance) may play a relevant role to modify the expected angular frequency. Obviously, the same consideration holds true for the amplitude variable.

We close this section with a final remark. One may argue that the drift terms $\hat a_{\theta_i}$ and $\hat \ba_{\bR_i}$ are artifacts due to It\^o interpretation. However, it turns out that the frequency drift is also present if Stratonovich interpretation is used \cite{yoshimura2008,bonnin2013,bonnin2014}. To clarify the point, consider the Stratonovich SDE describing the network of oscillators
\begin{equation}
d \bX_i = \left[ \ba_i(\bX_i) + \varepsilon \bc_i(\bX_1,\ldots, \bX_N) \right] dt + \varepsilon \bB_i(\bX_1, \ldots, \bX_N) \circ d\bW_i \qquad i=1,\ldots,N \label{eq27}
\end{equation}
Taking into account that in Stratonovich interpretation traditional calculus rules apply, repeating the procedure used in the previous section the following reduced phase model is derived
\begin{equation}
d\theta_i = \left[1 + \varepsilon c_{\theta_i}(\theta_1,\ldots,\theta_N) \right] dt + \varepsilon \bB_{\theta_i}(\theta_1,\ldots,\theta_N) \circ d\bW_i \label{eq28}
\end{equation}
However, in Stratonovich interpretation is no longer true that the stochastic processes and the noise increments are uncorrelated. Because of the anticipating nature discussed in section \ref{noisy network}, $E[\int_{t_0}^t f(\bX) \circ d\bW] \ne 0$. To resolve the correlation, a Stratonovich SDE has to be transformed into its equivalent It\^o SDE by the addition of the drift correction term \cite{oksendal2003,gardiner1985}. Here is where the drift coefficient, that arises naturally from the quadratic terms in It\^o formula, comes into play \cite{yoshimura2008,bonnin2013,bonnin2014}.

\section{Example}\label{example}

In this section we give an example to show the derivation of the phase and amplitude deviation equations starting from the network's state equations. To keep everything reasonably simple, we consider a network composed by two oscillators written in polar coordinates
\begin{subequations}
\begin{align}
d \rho_i & = \rho_i \left( 1-\rho_i \right) dt + \varepsilon \, \rho_i d W_{\rho_i} \label{eq29a} \\[1ex]
d \phi_i &  = \left[ \nu_i \rho_i + \varepsilon (\phi_j - \phi_i) \right]dt + \varepsilon \rho_j \, d W_{\phi_i} \label{eq29b}
\end{align}
\end{subequations}
for $i,j=1,2$, and $j\ne i$. The real parameters $\nu_i$ define the free running frequencies of the oscillators in absence of noise. Although most of the information concerning expectation values and phase locking can be directly obtained from equations \eqref{eq29a}--\eqref{eq29b}, we first transform these equations into the equivalent amplitude and phase models using theorems \ref{theorem1} and \ref{theorem2} to show the application of the method.


For $\varepsilon =0$, equation \eqref{eq29a} admits two stationary solutions: $\rho_i=0$ that corresponds to an unstable equilibrium point, and $\rho_i=1$, that corresponds to an asymptotically stable limit cycle with angular frequency $d\phi_i/dt = \nu_i$. Without loss of generality, we shall assume $\nu_1>\nu_2$. To investigate the synchronization of the two oscillators we look at the phase difference $\psi = \phi_1 - \phi_2$. The oscillators are phase locked if the phase difference remains constant in time. If the two oscillators have different free running frequencies $\nu_1 \ne \nu_2$, the phase difference $\psi = (\nu_1-\nu_2)t$ grows unboundedly large (in absolute value) as the time passes. Conversely, if the coupling effect is taken into account but the noise influence is ignored, the phase difference evolves according to
\[ \dfrac{d\psi}{dt} = \nu_1 - \nu_2 - 2 \varepsilon \psi \]
Asymptotically the two oscillators become phase locked with phase difference
\begin{equation}
\psi_s = \frac{\nu_1 - \nu_2}{2\varepsilon} \label{eq30}
\end{equation}
Moreover, since $d\psi/dt>0$ for $\psi < \psi_s$ and $d\psi/dt<0$ for $\psi > \psi_s$, the phase locked state is asymptotically stable.

\subsection{Amplitude and phase equations using an orthogonal frame}

Each uncoupled oscillator of the network \eqref{eq29a}--\eqref{eq29b} admits the limit cycle
\begin{equation}
\bx_{S_i}(\rho_i,\phi_i) = \left[ \begin{array}{c}
1 \\
\nu_i\,t
\end{array} \right]
\end{equation}
It follows that the orthogonal basis is given by the tangent vector $\bu_{1_i}(t) = [0,1]^T$ and the orthogonal unit vector $\bu_{2_i}(t) = [1,0]^T$. Since the matrix $\bU_i(t) = [\bu_{1_i}(t), \bu_{2_i}(t)]$ is orthogonal, we have $\bv_{1_i}(t) = \bu_{1_i}(t)$ and $\bv_{2_i}(t) = \bu_{2_i}(t)$. The change of coordinates \eqref{eq12} implies $\rho_i = 1 + R_i$ and $\phi_i = \nu_i \theta_i$. Since the Jacobian of the transformation is $|D\bh_i(\theta_i,R_i)\big|_{R_i=0} = \nu_i$, the coordinate transformation holds for any value of the noise intensity. Using theorem \ref{theorem1} it is straightforward to derive the amplitude and phase equations
\begin{subequations}
\begin{align}
d \theta_i = & \left[ 1 + R_i + \varepsilon \left( \dfrac{\nu_j}{\nu_i} \theta_j - \theta_i \right) \right] dt + \varepsilon \dfrac{1+R_j}{\nu_i} W_{\phi_i} \label{eq32a}\\[2ex]
d R_i = & \left[ - R_i ( 1+ R_i) \right] dt + \varepsilon (1+R_i) d W_{\rho_i} \label{eq32b}
\end{align}
\end{subequations}
As it was expected from theorem \ref{theorem2}, in the phase equation the drift coefficient contains a linear term in $R_i$.

\subsection{Amplitude and phase equations using a Floquet frame}

The Jacobian matrix of \eqref{eq29a}, \eqref{eq29b} for $\varepsilon=0$ evaluated over the limit cycle for $\varepsilon=0$ is
\begin{equation}
\bA_i (\bx_{S_i}) = \left[ \begin{array}{cc}
-1 & 0\\
\nu_i & 0
\end{array} \right]  \label{eq33}
\end{equation}
with eigenvalues $\lambda_{1_i}=0$, $\lambda_{2_i} = -1$, for all $i$. The corresponding eigenvectors are the Floquet vectors $\bu_{1_i}(t) = [0,1]^T$ and $\bu_{2_i}(t)=[1,-\nu_i]^T$. Inverting the matrix $\bU_i(t)=[\bu_{1_i}(t),\bu_{2_i}(t)]$ we find the Floquet co--vectors $\bv_{1_i}(t)=[\nu_i,1]^T$ and $\bv_{2_i}(t)=[1,0]^T$. The relation between the old and the new coordinates is $\rho_i = 1 +R_i$ and $\phi_i =\nu_i (\theta_i - R_i)$. As before the Jacobian matrix $D\bh_i$ is regular on the whole plane $\theta_i,R_i$. Consequently the phase and amplitude equations in the new basis are
\begin{subequations}
\begin{align}
d\theta_i = & \left\{ 1 -R_i^2 + \varepsilon \left[\dfrac{\nu_j}{\nu_i} (\theta_j - R_j) - (\theta_i - R_i)\right] \right\} dt + \varepsilon \left[ \mu_i (1+R_i) d W_{\rho_i} + \dfrac{1+R_j}{\nu_i} dW_{\phi_i}  \right] \label{eq34a} \\[2ex]
d R_i = & - \left[ R_i \left( 1+ R_i \right)\right] dt + \varepsilon \mu_i (1+R_i) dW_{\rho_i} \label{eq34b}
\end{align}
\end{subequations}
Comparing \eqref{eq32a} to \eqref{eq34a} we observe that according to theorem \ref{theorem2}, the latter has a drift coefficient that starts with a quadratic term in $R_i$. The use of Floquet basis also emphasizes the role played by higher order moments, e.g. the variance, on the angular frequencies of the oscillators.

In this particular example $E[R_i]$ and $E[R_i^2]$ can be determined analytically, because equation \eqref{eq34b} depends on $R_i$ only. Thus we can write a one dimensional Fokker--Planck equation (see \cite{gardiner1985}) for the probability density function (PDF) of the amplitude variable
\begin{equation}
\dfrac{\partial p_i(R_i,t)}{\partial t} = -\dfrac{\partial}{\partial R_i} \left[ -R_i\left( 1+R_i \right) p_i(R_i,t) \right] + \dfrac{\varepsilon^2}{2} \dfrac{\partial^2}{\partial R_i^2} \left[(1+R_i)^2 p_i(R_i,t)\right]\label{eq35}
\end{equation}
In the limit $t\rightarrow +\infty$ it admits the stationary solution
\begin{equation}
\overline p_i(R_i) = \dfrac{\mathcal{N}}{(1+R_i)^2} \exp \left\{\frac{2}{\varepsilon^2} \left[\ln(1+R_i) - R_i\right]\right\}\label{eq36}
\end{equation}
where $\mathcal{N}$ is normalization constant that can be determined through the requirement $\int_{-\infty}^{+\infty} \overline p_i(R_i) \,dR_i = 1$. Using the stationary PDF we can compute the expectation value for an arbitrary function of the amplitude through
\begin{equation}
E[f(R_i)] = \int_{-\infty}^{+\infty} f(R_i) \, \overline p_i(R_i) \, dR_i\label{eq37}
\end{equation}

\begin{figure}
\centering
 \includegraphics[width=70mm]{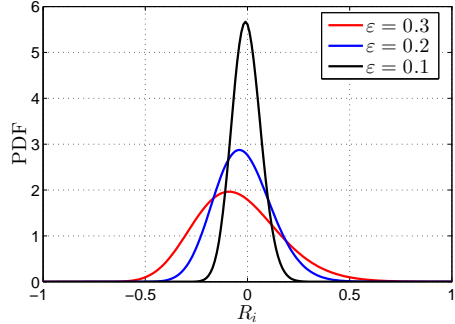}
\caption{Probability density function for the amplitude deviation given by \eqref{eq36}, for different values of the noise intensity $\varepsilon$.}\label{figure2}
 \end{figure}
\begin{figure}
\centering
 \includegraphics[width=60mm]{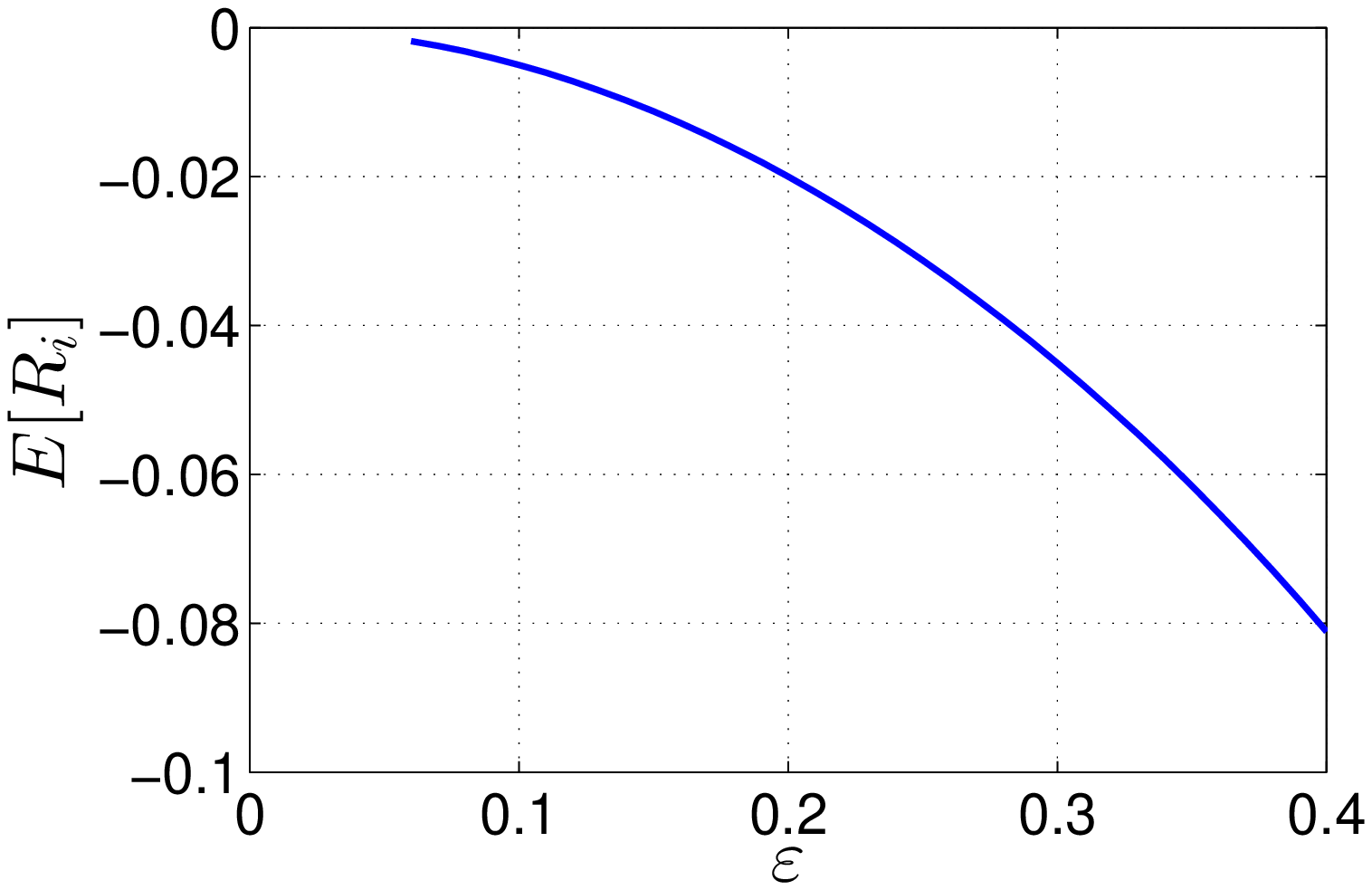}
 \includegraphics[width=60mm]{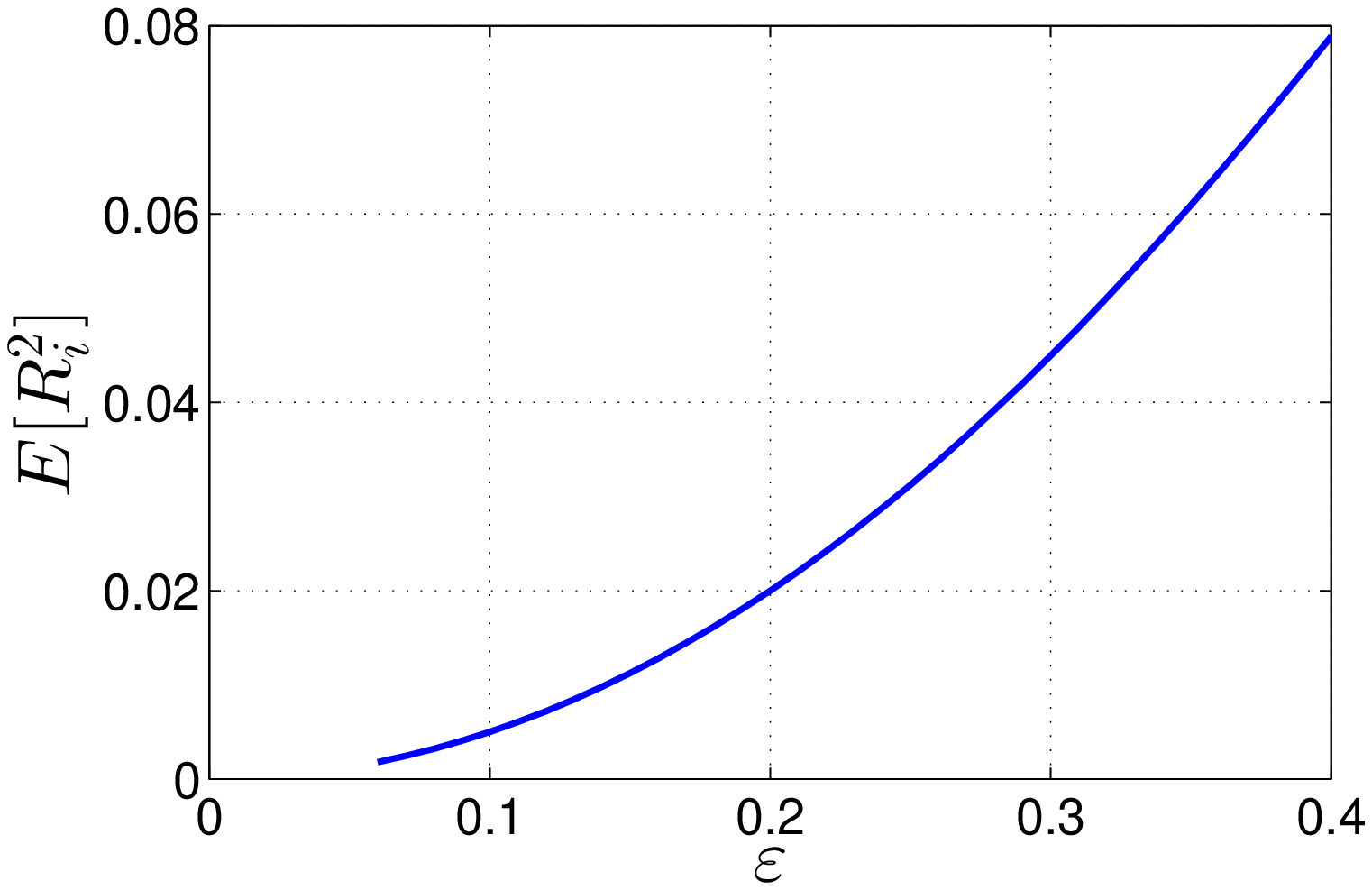}
\caption{Left: Expected amplitude $E[R_i]$ versus the noise intensity $\varepsilon$. Right: Expected squared amplitude $E[R_i^2]$ versus the noise intensity $\varepsilon$.}\label{figure3}
 \end{figure}

Figure \ref{figure2} shows the stationary PDF $\overline p_i(R_i)$ for different values of the parameter $\varepsilon$. For small noise intensity the PDF is well approximated by a Gaussian distribution. Increasing the noise intensity, we observe an increase in the variance (due to diffusion) and a shift in the mode (the most probable value of the amplitude). We also observe the PDF becomes asymmetric with respect to the maximum value, thus indicating that higher order moments become more and more relevant. Figure \ref{figure3} shows the first two moments for the amplitude deviation. It is seen the quadratic dependence on the noise intensity as predicted by theorem \ref{theorem1}.

\begin{figure}
\centering
 \includegraphics[width=60mm]{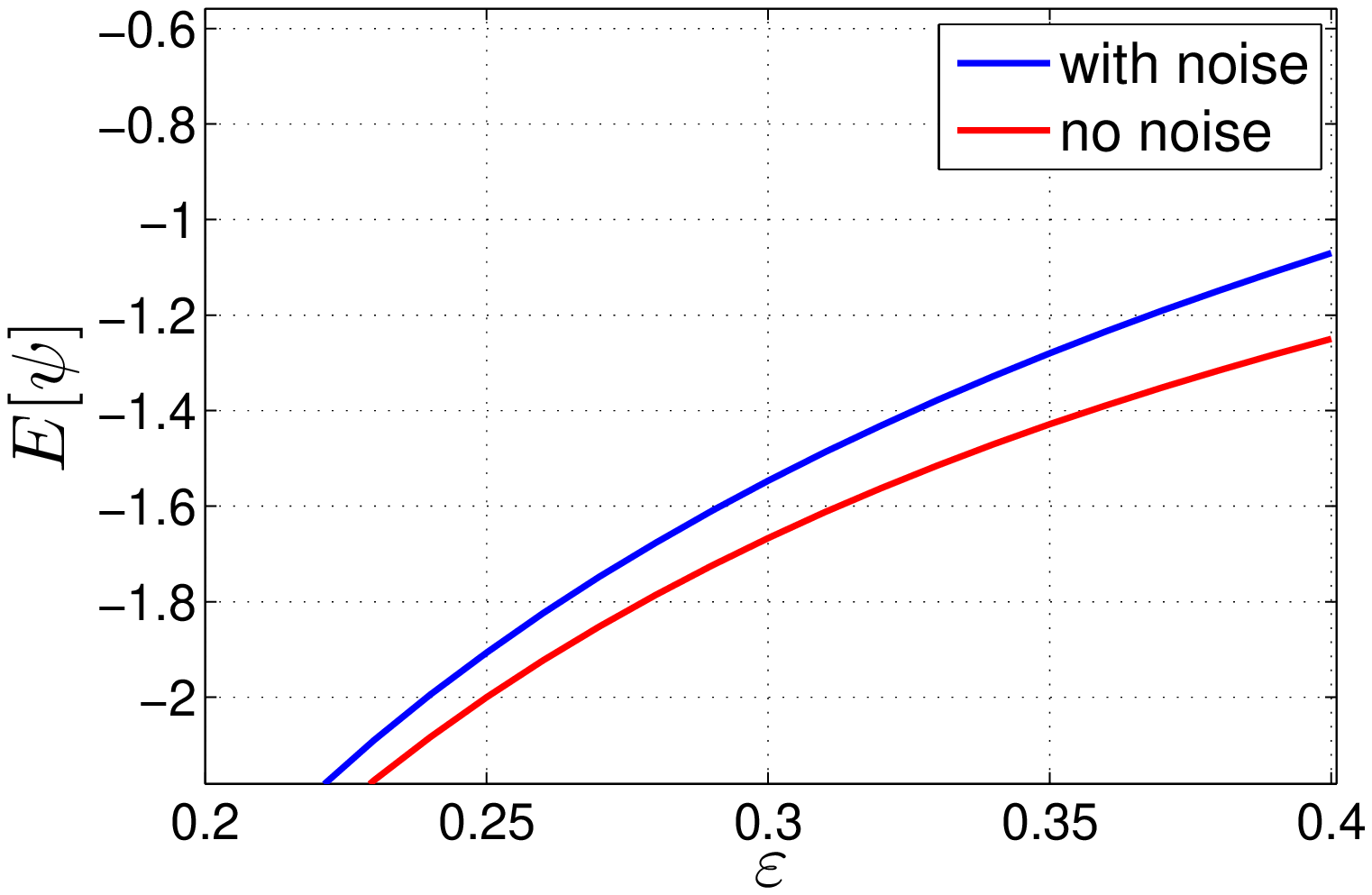}
  \includegraphics[width=60mm]{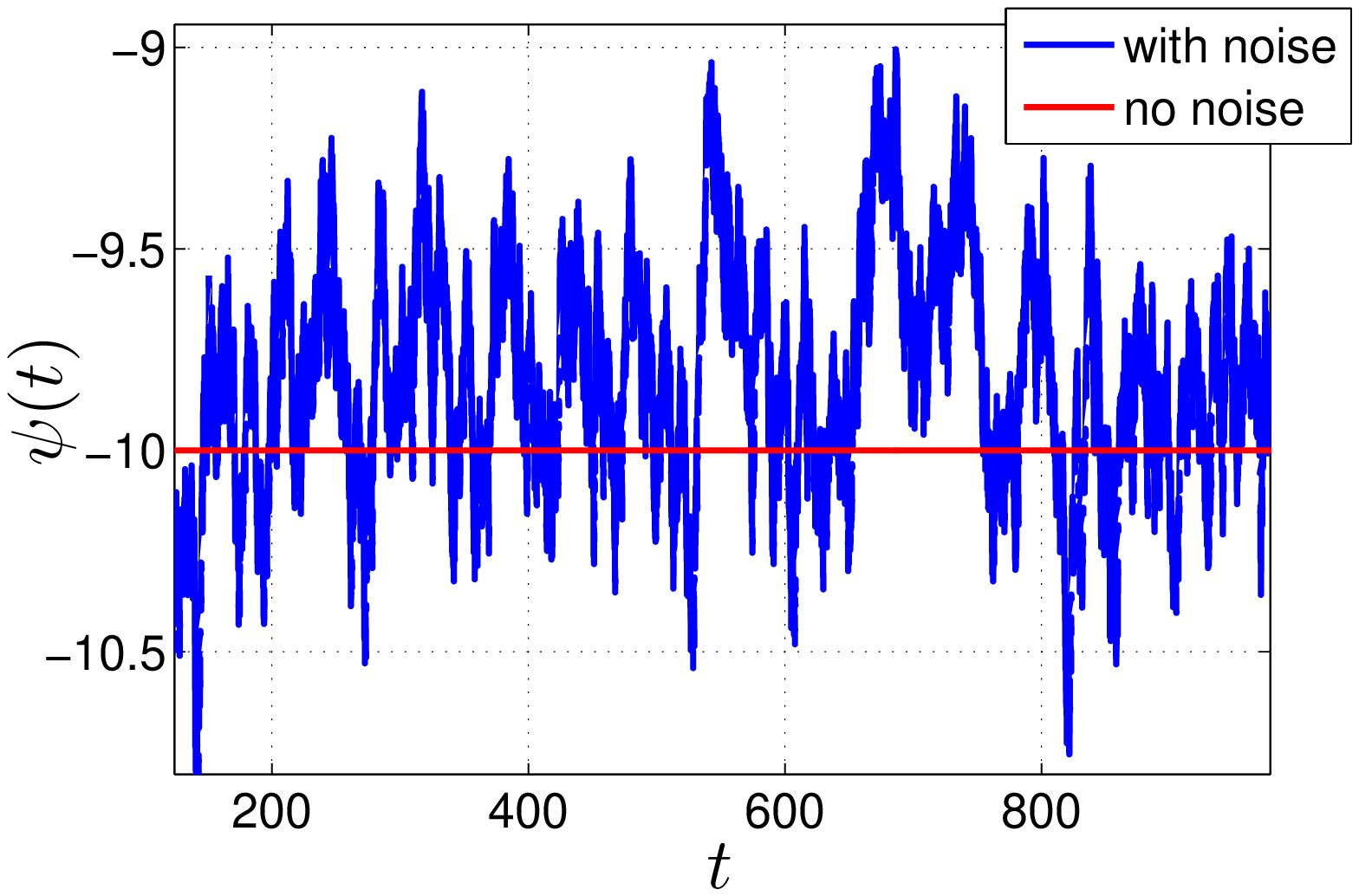}
\caption{Left: Comparison between the expected asymptotic phase difference $E[\psi]$ obtained solving \eqref{eq40} and the phase difference given by \eqref{eq30}, versus the noise intensity $\varepsilon$ for two oscillators with different free running angular frequencies, $\nu_1 = 1$ and $\nu_2 = 2$. Right: Phase difference for two oscillators (free running angular frequencies are $\nu_1 = 1$ and $\nu_2 = 2$ respectively), as a function of time for a specific realization of the noise. The noise intensity is set to $\varepsilon =0.05$. The phase difference in absence of noise is shown for reference.}\label{figure4}
 \end{figure}

Taking stochastic expectations of \eqref{eq34a} and using the zero expectation property of It\^o integral we have
\begin{align}
E\left[\dfrac{d\theta_i}{dt} \right] = & 1 - E\left[ R_i^2\right] + \varepsilon \left[\dfrac{\nu_j}{\nu_i} (E[\theta_j] - E[R_j]) - (E[\theta_i] - E[R_i])\right] \label{eq38}
\end{align}
Multiplying\footnote{We recall that $\theta_i$ is a normalized phase variable, multiplication for $\nu_i$ is necessary to retrieve the non normalized phase variable.}  by $\nu_i$ it is straightforward to obtain the equation for the expected phase difference
\begin{equation}
\dfrac{d E[\psi]}{dt} = \nu_i -\nu_j + \nu_j E[R_j^2] - \nu_i E[R_i^2]  - 2 \varepsilon \left(\nu_j E[R_j] - \nu_i E[R_i] \right) - 2 \varepsilon E[\psi]\label{eq39}
\end{equation}
Since the amplitude equation is the same for all the oscillators\footnote{The case where the noise intensity $\varepsilon$ is not equal for all the oscillators can be treated similarly. Obviously $E[R_i] \ne E[R_j]$ and $E[R_i^2] \ne E[R_j^2]$  and the solution of \eqref{eq39} would be more complicated.}, we have $E[R_1] = E[R_2]$ and $E[R_1^2] = E[R_2^2]$, then
\begin{equation}
\dfrac{d E[\psi]}{dt} = \left( \nu_1 -\nu_2 \right) \left( 1 - E[R_1^2] +2 \varepsilon E[R_1]\right) - 2 \varepsilon E[\psi]\label{eq40}
\end{equation}
Asymptotically the oscillators converge to the phase locked state
\begin{equation}
E[\psi] = \dfrac{\left(\nu_1-\nu_2 \right)\left( 1 - E[R_1^2] +2 \varepsilon E[R_1]\right) }{2\varepsilon}
\end{equation}
which is different from the phase locked state in absence of noise \eqref{eq30}. The phase difference in presence of noise is compared with that obtained without noise in figure \ref{figure4}. On the left we can see the asymptotic expected phase difference versus the noise intensity, while on the right it is shown the phase difference versus time for a specific realization of the noise process. It can be seen how noise operates to actively reduce the phase difference between the oscillators.

%

%

\section{Conclusions}\label{conclusions}

We have considered networks of coupled nonlinear oscillators subject to white Gaussian noise. We have shown that the network can be conveniently described by stochastic differential equations. The advantages and disadvantages of the two most popular interpretations, i.e. It\^o and Stratonovich, have been briefly outlined.

Using projection techniques and It\^o calculus, we have derived a rigorous mathematical description for the network dynamics, in terms of the phases and amplitude deviations of the oscillators. We have shown that using Floquet theory a partial decoupling between the phase and the amplitude dynamics can be obtained. This idea leads to the development of phase reduced models analogous to the celebrated Kuramoto model.

The amplitude and phase description highlights the influence of noise on the phases of the oscillators. It represents a good starting point for the analysis of the role of noise on synchronization. It is shown that noise can prevent phase locking of oscillators, by producing phase diffusion. It also shown that noise can favor synchronization. In fact the oscillators may adjust their frequency in response to noise intensity, and as a consequence noise can actively contribute to the synchronization by decreasing the frequency mismatch. A simple example has been used to illustrate the derivation of the amplitude and phase equations and their analysis.

\section*{Acknowledgement}
This work was supported by the \emph{Universit\`a Italo Francese} in the framework Galileo 2014--2015 (project number G14-145).

\section*{Appendix A: Proof of theorem \ref{theorem1}}\label{appendixA}
First we show that a neighborhood of the limit cycle exists, where $\theta_i$ and $\bR_i$ are It\^o processes. The Jacobian matrix of the coordinate transformation \eqref{eq12} evaluated on the limit cycle is
\[ D\bh_i(\theta_i,\bR_i)\big|_{\bR_i=0} = \big[r_i(\theta_i)\bu_{1_i}(\theta_i), \bu_{2_i}(\theta_i),\ldots, \bu_{n_i}(\theta_i) \big] \]
Since $\{\bu_{1_i}(t),\ldots,\bu_{n_i}(t)\}$ is a basis for $\R^{n}$, it follows that $\det D\bh_i(\theta_i,\bR_i)\big|_{\bR_i=0} \ne 0$. Then by the inverse function theorem a neighborhood of $\bR_i=0$ exists, where $\bh_i$ is invertible. Moreover, if $\bh_i$ is of class $\mathcal{C}^k$ then its inverse is also of class $\mathcal{C}^k$. Taking the inverse of $\bh_i$ we can write $\theta_i = \theta_i(\bx_i)$ and $\bR_i = \bR(\bx_i)$, and if the basis vectors are smooth enough it follows from  It\^o formula that $\theta_i$ and $\bR_i$ are It\^o processes.\\
Now we prove that $\theta_i$ and $\bR_i$ satisfy equations \eqref{eq13a} and \eqref{eq13b}. Using It\^o formula and eq. \eqref{eq9}, $\bx_i = \bh_i(\theta_i,\bR_i)$ implies

\begin{align*}
d\bx_i = &\frac{\partial \bh_i}{\partial \theta_i} d\theta_i + \frac{\partial \bh_i}{\partial \bR_i} d\bR_i +  \frac{1}{2} \, \frac{\partial^2 \bh_i}{\partial \theta_i^2} (d\theta_i)^2 +  \frac{1}{2} \, d\bR_i^T \frac{\partial^2 \bh_i}{\partial \bR_i^2} d\bR_i + \frac{\partial^2 \bh_i}{\partial \theta_i \partial \bR_i} d\theta_i d\bR_i \\
= & \big[ \ba_i(\bh_i(\theta_i,\bR_i)) + \varepsilon \bc_i(\bh_1(\theta_1,\bR_1),\ldots, \bh_N(\theta_N,\bR_N)) \big] dt \\
& + \bB_i(\bh_1(\theta_1,\bR_1),\ldots,\bh_N(\theta_N,\bR_N)) \, d\bW_i
\end{align*}
where $\partial \bh_i/\partial \bR_i$ and $\partial^2 \bh_i/\partial \bR_i^2$ are the Jacobian and the Hessian matrices, respectively. Introducing \eqref{eq12} yields
\begin{equation}\begin{array}{c}
\bigg( \ba_i(\bx_{S_i}) + \dfrac{\partial \bY_i}{\partial \theta_i} \, \bR_i \bigg) d \theta_i + \bY_i d\bR_i + \dfrac{1}{2} \bigg( \dfrac{\partial \ba_i(\bx_{S_i})}{\partial \theta_i} + \dfrac{\partial^2 \bY_i}{\partial \theta_i^2} \bR_i \bigg) (d\theta_i)^2
+ \dfrac{\partial \bY_i}{\partial \theta_i} \, d\theta_i \, d\bR_i \\[2ex]
= \big[ \ba_i(\bx_{S_i} + \bY_i \bR_i) + \varepsilon  \bc_i(\bx_{S_1} + \bY_1 \bR_1,\ldots,\bx_{S_N} + \bY_N \bR_N) \big] dt \\[2ex]
+ \varepsilon \bB_i(\bx_{S_1} +\bY_1 \bR_1,\ldots,\bx_{S_N} +\bY_N \bR_N) d\bW_i\label{eq15}
\end{array} \end{equation}
Multiplying \eqref{eq15} to the left by $\bv^T_{1_i}$ and using the bi--orthogonality condition we get
\begin{equation}\begin{array}{c}
\bigg( r_i + \bv_{1_i}^T \dfrac{\partial \bY_i}{\partial \theta_i} \, \bR_i \bigg) d \theta_i + \dfrac{1}{2} \bv_{1_i}^T \bigg( \dfrac{\partial \ba_i(\bx_{S_i})}{\partial \theta_i} + \dfrac{\partial^2 \bY_i}{\partial \theta_i^2} \bR_i \bigg) (d\theta_i)^2 + \bv_{1_i}^T \dfrac{\partial \bY_i}{\partial \theta_i} \, d\theta_i \, d\bR_i \\[2ex]
= \bv_{1_i}^T \left[\ba_i(\bx_{S_i} + \bY_i \bR_i) + \varepsilon \bc_i(\bx_{S_1} + \bY_1 \bR_1,\ldots,\bx_{S_1} + \bY_1 \bR_1) \right]dt \\[2ex]
+ \varepsilon \, \bv_{1_i}^T \bB_i(\bx_{S_1}  +\bY_1 \bR_1,\ldots,\bx_{S_N} + \bY_N \bR_N) d\bW_i \label{eq16}
\end{array} \end{equation}
Multiplying \eqref{eq15} to the left by $\bZ_i^T$ gives
\begin{equation}\begin{array}{c}
\bZ_i^T \dfrac{\partial \bY_i}{\partial \theta_i} \bR_i \, d\theta_i + d\bR_i + \dfrac{1}{2} \bZ_i^T \bigg( \dfrac{\partial \ba_i(\bx_{S_i})}{\partial \theta_i} + \dfrac{\partial^2 \bY_i}{\partial \theta_i^2} \bR \bigg) (d\theta_i)^2  + \bZ_i^T \dfrac{\partial \bY_i}{\partial \theta_i} \, d\theta_i \, d\bR_i \\[2ex]
= \bZ_i^T \left[ \ba_i(\bx_{S_i} + \bY_i \bR_i) + \varepsilon \bc_i(\bx_{S_1}  +\bY_1 \bR_1,\ldots,\bx_{S_N} + \bY_N \bR_N)  \right] dt \\[2ex]
+ \varepsilon \bZ_i^T \bB_i(\bx_{S_1} +\bY_1 \bR1,\ldots,\bx_{S_N} + \bY_N \bR_N) d\bW_i \label{eq17}
\end{array} \end{equation}
Since $\theta_i$ and $\bR_i$ are It\^o processes they satisfy relations of type $d \theta_i = \alpha_i \, dt + \boldsymbol{\beta_i} \, d\bW_i$, and $d \bR_i = \boldsymbol{\gamma}_i \, dt + \boldsymbol{\sigma}_i \, d\bW_i$, respectively. By It\^o lemma $(d\theta_i)^2 = \boldsymbol{\beta}_i \boldsymbol{\beta}_i^T \, dt$, and
$d\theta_i \, d\bR_i = \boldsymbol{\sigma}_i \boldsymbol{\beta}_i^T \, dt$. Introducing these results in \eqref{eq16}, \eqref{eq17} and equating terms in $d\bW_i$ we obtain
\begin{subequations}
\begin{align}
\boldsymbol{\beta}_i & =  \varepsilon \left( r_i + \bv_{1_i}^T \frac{\partial \bY_i}{\partial \theta_i} \bR_i \right)^{-1} \bv_{1_I}^T \, \bB_i(\bx_{S_1} + \bY_1 \, \bR_1,\ldots,\bx_{S_N} + \bY_N \, \bR_N) \label{eq18a}\\
\boldsymbol \sigma_i & =  \varepsilon \bZ_i^T \bB_i (\bx_{S_1} + \bY_1 \, \bR_1,\ldots,\bx_{S_N} + \bY_N \, \bR_N) - \bZ_i^T \frac{\partial \bY_i}{\partial \theta_i} \, \bR_i \,  \boldsymbol{\beta}_i \label{eq18b}
\end{align}
\end{subequations}
Finally, using \eqref{eq18a}, \eqref{eq18b} together with $(d\theta_i)^2 = \boldsymbol{\beta}_i \boldsymbol{\beta}_i^T \, dt$, $d\theta_i \, d\bR_i = \boldsymbol{\sigma}_i \boldsymbol{\beta}_i^T \, dt$ in \eqref{eq16}, $\eqref{eq17}$, and rearranging the terms we get the thesis. \hfill $\square$\\

\section*{Appendix B: Proof of theorem \ref{theorem2}}\label{appendixB}

First of all, we observe that by hypothesis the columns of matrix $\bP_i(t)$ are linearly independent for any $t$, and therefore they can be chosen as a basis for $\R^{n}$. Moreover, from equation \eqref{eq11}, we have
$r_i\bu_{1_i}(t)=d\bx_{S_i}/dt$. Therefore $\nu_{1_i}=0$, since $d\bx_{S_i}/dt$ is the solution of the variational equation associated to the structural Floquet exponent. Furthermore, equation \eqref{eq19} implies $\bP_i(t)=\Phi_i(t) \bP_i(0)e^{-\bD_i t}.$ Taking the derivatives of \eqref{eq19} yields
\[ \dfrac{d\Phi_i}{dt} = \dfrac{d\bP_i}{dt} \, e^{\bD t}\bP_i^{-1}(0)-\bP_i(t)\bD_i e^{\bD_i t}\bP_i^{-1}(0)\]
and taking into account that $\Phi_i(t)$ is a fundamental matrix of the variational equation we obtain:
$d\bP_i/dt=\bA_i \bP_i-\bP_i\bD_i$
and consequently
$d\bY_i/dt=\bA_i \bY_i-\bY_i\widetilde \bD_i.$
Substituting the expression of $d\bY_i/dt$ in \eqref{eq14a}, \eqref{eq14e} and \eqref{eq14f}, taking the Taylor series $\ba_i(\bx_{S_i}+\bY_i\bR_i)=\ba_i(\bx_{S_i})+\bA_i\bY_i\bR_i+\dots$, and using the bi-orthogonality condition the thesis follows. \hfill $\square$\\

\end{document}